\documentclass[aps,prl,reprint,superscriptaddress]{revtex4-2}

\usepackage{graphicx}
\usepackage{microtype}
\usepackage{xcolor}
\usepackage{nicefrac}
\usepackage{microtype}
\usepackage{nicefrac}
\usepackage{amsmath}
\usepackage{amssymb}
\usepackage{gensymb}
\usepackage{ulem}
\usepackage{stix}
\usepackage[T1]{fontenc}
\usepackage[colorlinks=true, linkcolor=blue, anchorcolor=black, citecolor=blue, filecolor=black, urlcolor=blue]{hyperref}
\usepackage[all]{hypcap}

\newcommand{\ikf}{Institut f\"ur Kernphysik, Goethe-Universit\"at Frankfurt, Max-von-Laue-Str. 1, 60438 Frankfurt am Main, Germany}
\newcommand{\ESRF}{ESRF, The European Synchrotron, 71 Avenue des Martyrs, CS 40220, 38043 Grenoble Cedex 9, France}

\newcommand{\berlin}{Molecular Physics, Fritz-Haber-Institut der Max-Planck-Gesellschaft, Faradayweg 4-6, 14195 Berlin, Germany}
\newcommand{\HD}{Max-Planck-Institut f\"{u}r Kernphysik, Saupfercheckweg 1, 69117 Heidelberg, Germany}
\newcommand{\EXFEL}{European XFEL, Holzkoppel 4, 22869 Schenefeld, Germany}
\newcommand{\Paris}{Sorbonne Université, CNRS, Laboratoire de Chimie Physique Matière et Rayonnement, UMR 7614, F-75005 Paris, France}

\begin{document}
\title {
\large
Selective Bond Breaking in CO$_2^{2+}$ Induced by Photoelectron Recoil} 
\author{J.~Weiherer} \address{\ikf}
\author{N.~Melzer}  \address{\ikf}
\author{M.~Kircher} \address{\ikf}
\author{A.~Pier} \address{\ikf}
\author{L.~Kaiser} \address{\ikf}
\author{J.~Kruse} \address{\ikf}
\author{N.~Anders} \address{\ikf}
\author{J.~Stindl} \address{\ikf}
\author{L.~Sommerlad} \address{\ikf}
\author{O.~D.~McGinnis} \address{\ikf}
\author{M.~Schmidt} \address{\ikf}
\author{J. Drnec} \address{\ESRF}
\author{F.~Trinter} \address{\berlin}
\author{M.~S.~Sch\"offler} \address{\ikf}
\author{L.~Ph.~H.~Schmidt}\address{\ikf}
\author{N.~Sisourat} \address{\Paris}
\author{S.~Eckart}\address{\ikf}
\author{T.~Jahnke} \address{\HD}\address{\EXFEL}
\author{R.~D\"orner} \email{doerner@atom.uni-frankfurt.de}
\address{\ikf}

\begin{abstract}
After core-ionization of CO$_2$, typically an Auger-Meitner decay takes place, leading to the formation of a dicationic molecule that may dissociate into CO$^+$ and O$^+$. We demonstrate experimentally that the recoil momentum of the photoelectron steers, which of the two equivalent bonds breaks during the dissociation. At 20~keV photon energy, we observe an asymmetry of up to 25\% for bond cleavage that depends on the emission direction of  the photoelectron. Furthermore, we show that this effect leads to a significant nondipole effect in molecular dissociation in the laboratory frame: O$^+$ fragments are more likely to be emitted in the direction opposite to the light propagation than along it. 

\end{abstract}
\maketitle

The inherent symmetries of molecules dictate their fundamental properties. In diatomic homonuclear molecules or triatomic linear molecules of the type ``A-B-A'', the molecules' eigenstates have a well-defined parity, leading to characteristic rotational spectra and to a symmetric bond dissociation probability. Achieving symmetry breaking in isolated molecules requires the preparation of a superposition state involving electronic or vibrational states of opposite parity. Electronic-state superposition has been the subject of extensive experimental investigations. Such Raman-type processes have enabled indirect excitation of nuclear degrees of freedom through electronic interactions; however, a direct manipulation of nuclear degrees of freedom to induce parity breaking has yet to be realized. Here, we report an experiment demonstrating that already the fundamental process of photoionization induces (by concept) a molecular symmetry breaking through photoelectron recoil-induced nuclear dynamics.   

Symmetry breaking through control of electronic degrees of freedom has been demonstrated along a breadth of different schemes, including symmetry-broken laser pulses, measurement-induced asymmetries, and schemes exploiting correlations or entanglement of photoelectrons and dissociating molecular ions. In a pioneering experiment, Kling et al. \cite{Kling.2006} broke the symmetry of H$_2^+$ dissociation using a carrier-envelope-stabilized ultra-short laser pulse. Later, phase-locked two-color pulses \cite{Betsch.2010,Ray.2009} have been shown to be highly effective in breaking the electronic state symmetry. Such pulses have also been used to control the bond breaking in CO$_2$ \cite{Betsch.2010,Endo.2017}. Continuing in a similar vein, the detection of the electron emission direction can be used to break the symmetry of the laser-driven dissociation process \cite{Wu.2013}.
A fascinating alternative, which does not require symmetry-broken external fields, is the use of correlations or entanglement between the direction of an emitted electron and the left-behind molecular ions. Along this line, Serov and Kheifets \cite{Serov.2014} have suggested that the field of a slow photoelectron can mix the gerade and ungerade electronic states of H$_2^+$, leading to asymmetric dissociation, an approach realized experimentally by Waitz et al. later on \cite{Waitz.2016}. Entanglement between electron and dissociating molecular ions as a source of electronic symmetry breaking was demonstrated by Martín et al. \cite{Martin2007} for dissociative ionization through doubly excited states of H$_2$, and asymmetries have also been found for the dissociation of CO$_2$ and CS$_2$ in single photoionization \cite{Guillemin.2015,Liu.2008,Miyabe.2009,Sturm2009}.  Although very different in detail, most of these approaches have in common that they control the electronic part of the molecular wave function. The possibly symmetry-broken nuclear motion (e.g., in the dissociation of CO$_2$) is only an indirect consequence.

A route to directly drive nuclear degrees of freedom, bypassing the need to change the potential energy landscape by electronic excitations, was already proposed by Domcke and Cederbaum \cite{Domcke.1978} in 1978. They suggested that the momentum transfer in photoionization can directly induce motion of the nuclei inside a molecule. They showed that, as a consequence of momentum conservation, the momentum of a photoelectron or an Auger electron is compensated by the recoil momentum of the respective nucleus. This momentum transfer directly leads to vibrational and rotational excitation \cite{Kukk.2005,Thomas.2011,Miron.2014,Kukk.2018}. Such manipulation of nuclear degrees of freedom can then, as a secondary effect, modify the electronic degrees of freedom, such as changing electronic decay processes (for example, such as ICD \cite{Kreidi.2009}) and driving nonadiabatic effects of molecular dynamics \cite{Cederbaum.2009}. Ultimately, this momentum transfer could even influence bond breaking as predicted by Liu et al. \cite{Liu2019}. This prediction is experimentally verified in this Letter.

In the present work, we investigate the role of the photoelectron recoil on the dissociation of $\text{CO}_2^{2+}\rightarrow \text{CO}^+ + \text{O}^+$ initiated by O 1s or C 1s photoionization, followed by Auger-Meitner decay. At a photon energy of $E_\gamma=20$~keV, the momentum transfer of the photoelectron is 38~a.u., which corresponds to an energy of 
0.90~eV (0.67~eV) if imparted to a C (O) atom. As we show in this Letter, this recoil significantly influences the dissociation dynamics, selectively cleaving one CO bond or the other. 

The momentum transfer $\vec{Q}$ to the nucleus is calculated by subtracting the momenta of the photoelectron $\vec{k}_e$ (38~a.u.) and the Auger electron $\vec{k}_{e_A}$ (4.6~a.u. for C 1s or 6.3~a.u. for O 1s) from the photon momentum $\vec{k}_\gamma$ (5.4~a.u.). 
In our experiment, the photoelectron momentum was not measured directly, but it can be inferred by subtracting the photon momentum and the measured Auger electron momentum from the sum momentum of the two ionic fragments. In more detail, the experiment was performed at beamline ID31 at the European Synchrotron Radiation Facility (ESRF) in Grenoble, France, using a COLTRIMS reaction microscope \cite{doerner00,outUllrich2003, Jahnke.2004} to measure the momentum vectors of the Auger electron and the molecular fragments CO$^+$ and O$^+$ in coincidence. The experimental setup was the same as in Ref. \cite{Melzer.2024}. Photons with a bandwidth of about 2\% were selected from the undulator beam using a pinhole monochromator \cite{outVaughankv5084}. The resulting photon beam with an intensity of around $2.6\times10^{14}$~photons/s was crossed with a supersonic CO$_2$ gas-jet target.
All fragment ions and electrons with an energy below $2$~keV were guided by parallel $50$~V/cm electric and $31$~G magnetic fields onto two time- and position-sensitive detectors with hexagonal delay-line readout \cite{jagutzki2002}. The ion arm of the spectrometer had a total length of $840.4$~mm and included an electrostatic lens \cite{DORNER1997225} to obtain the required ion momentum resolution of about $0.7$~a.u. The Auger electrons were measured directly; as indicated above, the photoelectron momentum was obtained by exploiting momentum conservation. Events in which Compton scattering induced double ionization can be separated from those created by photoionization due to their much smaller sum momentum of the two ionic fragments (see Fig.~1 in Ref. \cite{Schmidt.2024}). We achieved a triple coincidence (Auger electron, CO$^+$, O$^+$) rate of about $6$~Hz. 

\begin{figure}[!t]
\includegraphics{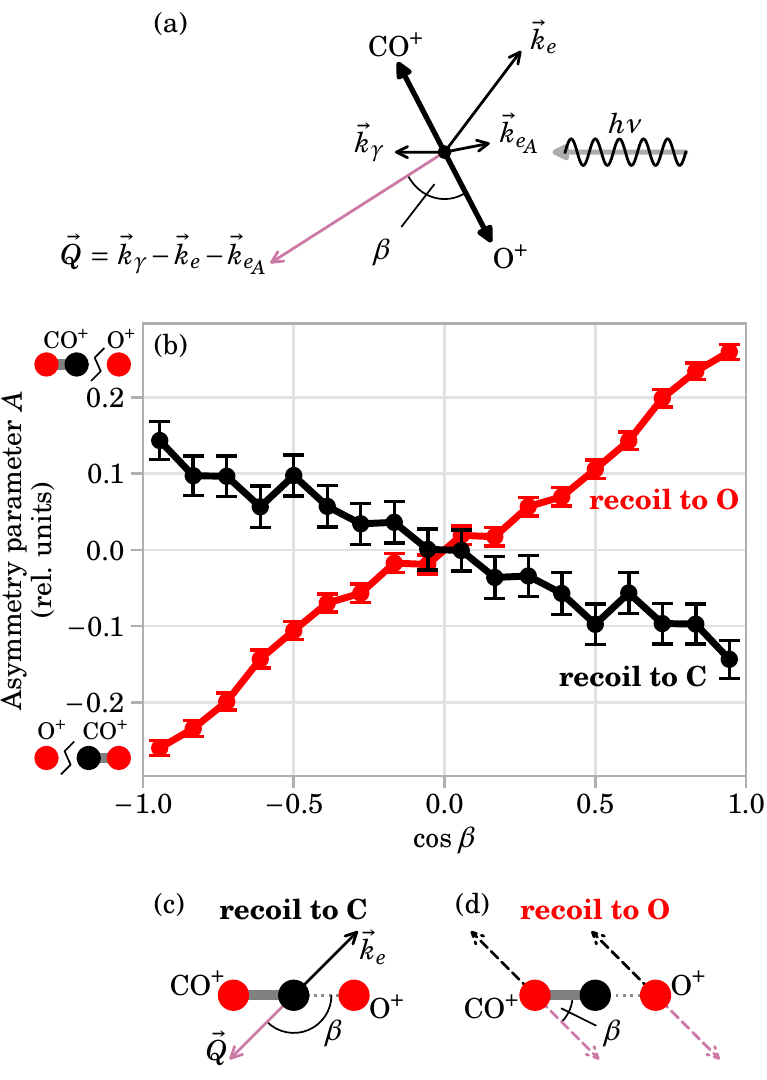}
\caption{Bond breaking asymmetry. (a) Sketch of the definition of angle $\beta$, which corresponds to the angle between the recoil-induced momentum transfer and the breakup direction of the molecule. Note that the magnitude of $\vec{k}_e$ is in reality much larger than depicted in the sketch. (b) Recoil-induced asymmetry in bond breaking for $\text{CO}_2^{2+}\rightarrow \text{CO}^+ + \text{O}^+$ induced by O 1s (red) and C 1s (black) photoionization by 20~keV photons, see text for details. Horizontal axis: cosine of the angle $\beta$ (see panel a). Vertical axis: The asymmetry parameter $A(\cos\beta)$ characterizes the difference in probability of breaking either of the two CO bonds as indicated by the pictograms. The error bars correspond to the standard statistical error. (c) Schematic view of the most preferred bond-breaking following C 1s ionization. For simplicity, the comparably small momenta of the Auger electron and the photon are omitted. (d) Same as (c) for O 1s ionization, see text for details.}
\label{fig1}
\end{figure}

Figure~\ref{fig1} shows the key result of our study. We plot the asymmetry parameter $A$ of the bond-breakage probability as a function of $\cos\beta$, where $\beta$ is the angle between the momentum transfer $\vec{Q}$ onto the atomic core from which the $K$-shell photoelectron is emitted and the fragmentation direction [see Fig.~\ref{fig1}(a)]. The latter is given by the vector 
$\vec{k}_\mathrm{rel}=\nicefrac{1}{2}\big(\vec{k}_{\text{O}^+} -\vec{k}_{\text{CO}^+}\big)$ which points from the $\text{CO}^+$ to the $\text{O}^+$ fragment. 

The asymmetry parameter $A$ is computed by taking the normalized difference of the number $N$ of measured CO$^+/$O$^+$ events for $\cos\beta$ and $-\cos\beta$, that is,
\begin{equation*}
    A(\cos\beta) = \frac{N(\cos\beta)-N(-\cos\beta)}{N(\cos\beta)+N(-\cos\beta)} \ ,
\end{equation*}
where, as explained above, $\beta$ is given relative to the emission direction of the O$^+$ fragment.

If either of the two CO bonds break with equal probability, the number of events $N$ at a given $\cos\beta$ would be independent of the sign of $\cos\beta$, i.e. $N(\cos\beta)=N(-\cos\beta)$.
As Fig.~\ref{fig1}(b) shows, this is not the case and the direction of the recoil-induced momentum transfer determines which bond breaks in the molecule. The figure shows two cases that we can distinguish: By inspecting the measured Auger electron energy, we can determine whether the O 1s or the C 1s shell was ionized in our experiment. We will discuss the C 1s ionization first. Figure~\ref{fig1}(c) shows the situation for negative values of $\cos\beta$. In this case, the right CO bond is stretched and the left bond is compressed by the momentum transfer, and, as the corresponding positive values of $A$ for $\cos\beta<0$ show, the stretched bond is more likely to break than the compressed one. In terms of symmetry breaking, the momentum transfer will excite a coherent superposition of the symmetric ($g$) and asymmetric ($u$) stretching modes of the molecule. The relative phase between these two excitations determines the probability of breaking the left or the right bond. This is the nuclear wave function analogue to the heavily discussed control of electronic localization in $\text{H}_2^+ \rightarrow p^+ + \text{H}$ fragmentation by control of the relative phase between excitation of $g$ and $u$ electronic states \cite{Kling.2006}.

The case of O 1s ionization is slightly more involved [Fig.~\ref{fig1}(d)]. A momentum transfer toward the right imparted onto the right O stretches the right bond, making it more likely to break. A momentum transfer in the same direction onto the left O compresses the left bond, which, as a consequence, also makes the right bond more likely to break. Therefore, for positive values of $\cos\beta$, as depicted in Fig.~\ref{fig1}(d), $A$ is positive.

Our results show that the asymmetry for O 1s photoionization is even more pronounced than that for C 1s photoionization.

\begin{figure}[!t]
\includegraphics{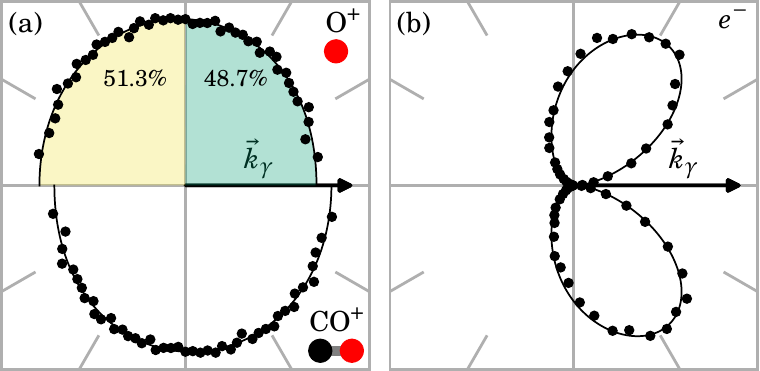}
\caption{Fragmentation asymmetries in the laboratory frame for O 1s photoionization of CO$_2$ by 20~keV photons. (a) Angular distribution of the ionic fragments relative momenta. The emission direction of the O$^+$ fragments relative to the propagation direction of the light $\vec{k}_{\gamma}$ are displayed in the upper half plane, that of the CO$^+$ fragments in the lower half plane. (b) Angular distributions of O 1s photoelectrons. In all panels, the photon propagation direction is from left to right. Error bars corresponding to the standard statistical error are smaller than the dot size. To guide the eye, the data in panels (a) and (b) were fitted with Legendre polynomials up to degree 3 and 9, respectively.}
\label{fig2}
\end{figure}

While the internal coordinate frame given by the electron momentum and $\vec{k}_\mathrm{rel}$ is appropriate for dissecting the underlying molecular dynamics and understanding the origin of symmetry breaking, its real-world consequences of the symmetry breaking (e.g., for randomly oriented CO$_2$ molecules in the Earth's atmosphere) would require the effect to be present in the laboratory frame, that is, to survive averaging over all electron emission directions.
Figure~\ref{fig2}(a) shows that this is the case. Upon dissociation of CO$_2^{2+}$, the CO$^+$ fragment is more likely to be emitted in the light propagation direction, while the O$^+$ fragment is more likely to be emitted backward, i.e., toward the photon source. This is impossible within the dipole approximation of light-matter interaction, in which the photon is only characterized by its polarization and not by its propagation direction. Figure~\ref{fig2}(b) illustrates the reason for the observed unidirectionality of the fragmentation. It shows the angular distribution of the photoelectrons from O 1s ionization. At photon energies of 20~keV, the angular distribution is a strongly forward-bent (in light propagation direction) two-lobe distribution. This strong nondipole effect on the electron emission entails a mirror image in the backward-directed recoil momentum (compare Ref. \cite{Grundmann2020}) which, in turn, via the recoil-driven symmetry-broken molecular dynamics (Fig.~\ref{fig1}), leads to the asymmetric dissociation in the laboratory frame shown in Fig.~\ref{fig2}(a). Note, this asymmetric emission of fragments from symmetric molecules in the laboratory system is not directly induced by the photon momentum. The photon momentum is of comparable magnitude to the Auger electron's recoil momentum. Both momenta are sufficient to cause rotational excitation, but both are too small to alter the dissociation dynamics significantly. However, although the comparatively small magnitude of the photon momentum $k_\gamma=E_\gamma/c$ does not directly cause asymmetric fragmentation, unlike the Auger electron recoil momentum, it has a dramatic amplification mechanism: it alters the photoelectron angular distribution, which, in turn, exerts a much larger jolt of $k_e \approx \sqrt{2m_e c k_\gamma}$ (neglecting the binding energy) onto the nucleus.

\begin{figure}[!t]
\includegraphics{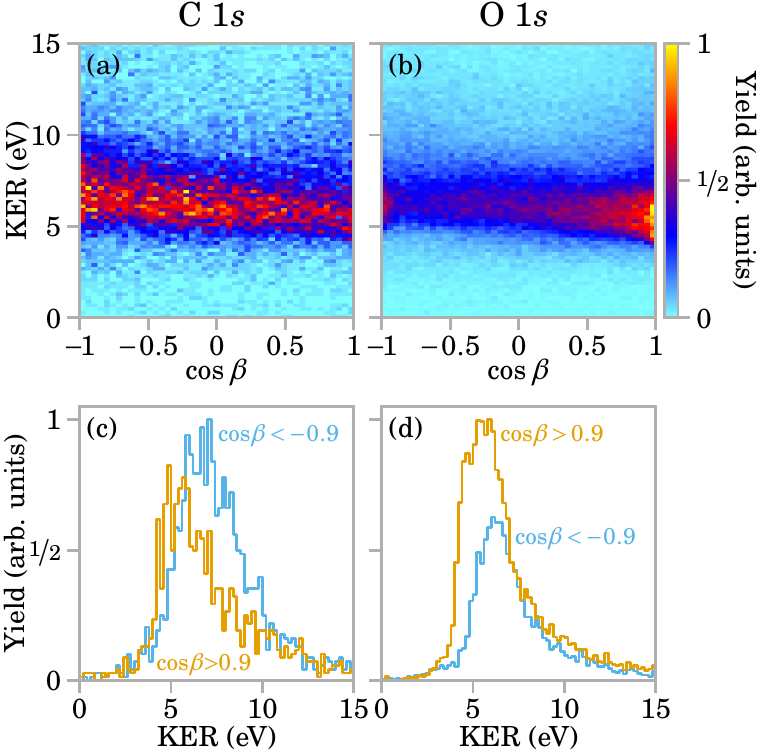}
\caption{KER distributions for C 1s (a,c) and O 1s (b,d) photoionization of CO$_2$ by 20~keV photons. (c,d) Projections onto the KER axis of panels (a,b) with $\cos{\beta}<-0.9$ and $\cos{\beta}>0.9$.}
\label{fig3}
\end{figure}

As a final aspect, we explore the photoelectron-recoil-induced dissociation in more detail by experimentally inspecting the ion kinetic energy. Figure~\ref{fig3} shows the kinetic energy release (KER) as a function of $\cos\beta$. The data for C 1s ionization [Figs.~\ref{fig3}(a,c)] and O 1s ionization [Figs.~\ref{fig3}(b,d)] show the enhanced yield for the breaking of the bond that was stretched by the recoil-induced momentum transfer. This is in line with the asymmetry parameter shown in Fig.~\ref{fig1}. In addition to this trend, we also find a significant change of the KER depending on the direction of the momentum transfer. The total energy of the CO$_2^{2+}$ intermediate state before the fragmentation is given by the potential energy surfaces populated by a Franck-Condon-type transition plus the kinetic energy given by the momentum transfer (0.90~eV and 0.67~eV for C 1s and O 1s, respectively). This initial energy can be expected to be approximately independent of $\cos\beta$. In the final state, this energy is distributed among the KER and internal vibrational and rotational excitation of the CO$^+$ fragment, which does depend on $\cos\beta$, resulting in the KER distributions of Fig.~\ref{fig3}.

In conclusion, we have shown that the recoil of the photoelectron can lead to selective molecular bond breaking. For symmetric molecules, this leads to a light-induced symmetry breaking, which is driven by directly exciting nuclear degrees of freedom rather than excitation of electronic degrees of freedom. At high photon energies, where the effect is substantial, strong nondipole effects on the electron emission direction create an asymmetry of the dissociation of CO$_2^{2+}$ in the laboratory frame as well. In interstellar clouds or in the atmosphere of the Earth, this leads to CO$^+$ fragments from irradiated CO$_2$ being preferentially ejected away from the radiation source (that is, into the atmosphere), while the O$^+$ fragments are ejected oppositely. We expect that the observed recoil-induced bond breaking is not limited to CO$_2$ and also occurs for more complex molecules. E.g. for two- and three-dimensional molecules, the effect will create chiral fragmentation patterns from achiral molecules without the need to address electronic degrees of freedom.

\begin{acknowledgments}
This work was funded by the Deutsche Forschungsgemeinschaft (DFG, German Research Foundation) -- Project No. 328961117 -- SFB 1319 ELCH (Extreme light for sensing and driving molecular chirality). The experimental setup was supported by BMBF. We acknowledge the European Synchrotron Radiation Facility (ESRF) for provision of synchrotron radiation facilities under proposal number CH-6729 (\href{https://doi.org/10.15151/ESRF-ES-1299363304              }{DOI: 10.15151/ESRF-ES-1299363304}) and we would like to thank V.~Honkimäki, H.~Isern, and F.~Russello for assistance and support in using beamline ID31. F.T. acknowledges funding by the Deutsche Forschungsgemeinschaft (DFG, German Research Foundation) -- Project 509471550, Emmy Noether Programme.
\end{acknowledgments}

\bibliography{main}
\bibliographystyle{apsrev4-2}

\end{document}